\documentclass[pra,superscriptaddress,groupedaddress,a4paper,twocolumn,nofootinbib]{revtex4}
\usepackage{amsmath,amsthm,amsfonts,amssymb,times,graphicx,color}
\usepackage{multirow}
\usepackage{subfigure}
\usepackage{mathtools}
\usepackage{url}
\usepackage{color}
\mathtoolsset{showonlyrefs}  

\newtheorem{theorem}{Theorem}[section]

\renewcommand{\hat}{}
\renewcommand{\top}{\intercal}

\newcommand{\couic}[1]{}
\newcommand{\couicfootnote}[1]{}
\newcommand{\couicefootnote}[1]{}

\newcommand{\ket}[1]{| #1 \rangle}

\begin{document}

\title{A quantum walk with both a continuous-time and a continuous-spacetime limit}

\author{Giuseppe Di Molfetta\footnote{Corresponding author.}}
\email{giuseppe.dimolfetta@lis-lab.fr}
\affiliation{Aix-Marseille Univ., CNRS, LIF, Marseille, France}

\author{Pablo Arrighi}
\email{pablo.arrighi@univ-amu.fr}
\affiliation{Aix-Marseille Univ, CNRS, LIS, and INRIA, ENS Paris-Saclay, LSV, Universit\'e Paris-Saclay, France}

\date{\today}
\begin{abstract}
\begin{center}
\bf{Abstract}
\end{center}
Nowadays, quantum simulation schemes come in two flavours. Either they are continuous-time discrete-space models (a.k.a Hamiltonian-based), pertaining to non-relativistic quantum mechanics. Or they are discrete-spacetime models (a.k.a Quantum Walks or Quantum Cellular Automata-based) enjoying a relativistic continuous spacetime limit. We provide a first example of a quantum simulation scheme that unifies both approaches. The proposed scheme supports both a continuous-time discrete-space limit, leading to lattice fermions, and a continuous-spacetime limit, leading to the Dirac equation. The transition between the two can be thought of as a general relativistic change of coordinates, pushed to an extreme. As an emergent by-product of this procedure, we obtain a Hamiltonian for lattice-fermions in curved spacetime with synchronous coordinates. 
\end{abstract}
\maketitle

\section{Introduction}

\begin{figure}
{\center
\includegraphics[scale=0.8]{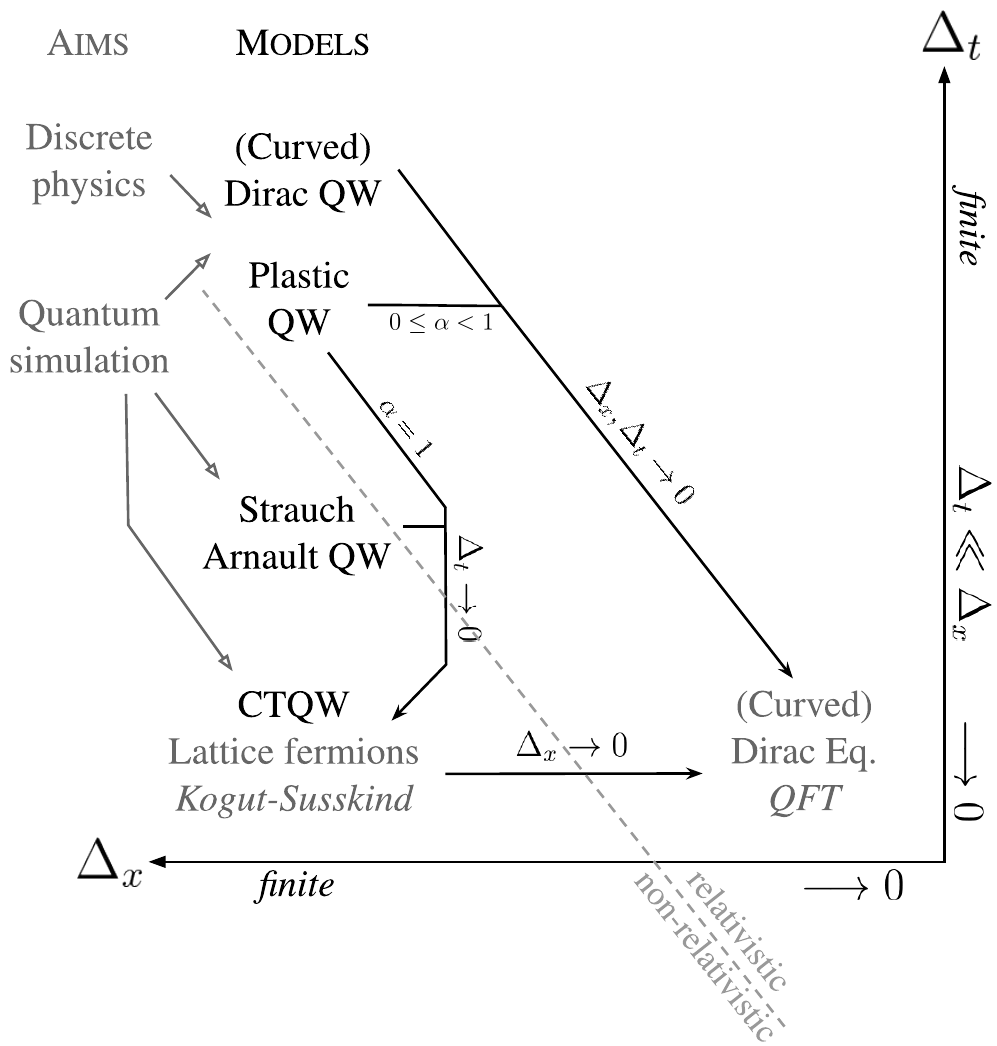}
}
\caption{{\em The Plastic QW} admits both a continuous-time discrete-space limit (to lattice fermions) and a continuous-spacetime limit (to the Dirac equation).}
\label{FigUnifiedQW}
\end{figure}

Confronted with the inefficiency of classical computers for simulating quantum particles, Feynman realized that one ought to use quantum computers instead \cite{FeynmanQC}. What better than a quantum system to simulate another quantum system? An important obstacle however is understand to which extent quantum systems can be represented by discrete quantum models. Indeed, whenever we simulate a physical system on a classical computer, we first need a discrete model of the physical system ``in terms of bits'': a simulation algorithm. Similarly, quantum simulation requires a discrete quantum model of the quantum system ``in terms of qubits'': a quantum simulation algorithm.

In the recent years, several such quantum simulation schemes have been devised \cite{JLP14,georgescu2014quantum, QuantumClassicalSim,HamiltonianBasedSchwinger}. Some of which were experimentally implemented \cite{BlattDirac}, including for interacting quantum particles \cite{InnsbruckLGT,ErezCiracLGT}. Most often these discrete models are Hamiltonian-based, meaning that they are discrete-space continuous-time ($\Delta_x\ \textrm{finite},\ \Delta_t\longrightarrow 0$). Their point of departure is always a discrete-space continuous-time reformulation of the target physical phenomena (e.g. the Kogut-Susskind Hamiltonian~\cite{kogut1975hamiltonian} formulation of quantum electrodynamics). Next, they either look for a quantum system in nature that mimics this Hamiltonian, or they perform a staggered trotterization of it in order to obtain unitaries \cite{StrauchCTQW,ArrighiChiral}. But even when the Hamiltonian is trotterized, time steps need remain orders of magnitude smaller than space steps ($\Delta_t\ll\Delta_x$). Thus, in either case, by having first discretized space alone and not time, Hamiltonian-based schemes take things back to the non-relativistic quantum mechanical setting: Lorentz-covariance is broken; the bounded speed of light can only be approximately recovered (via Lieb-Robinson bounds, with issues such as those pointed in \cite{EisertSupersonic,Osborne19}). This also creates more subtle problems such as fermion doubling, where spurious particles are created due to the periodic nature of the momentum space on a lattice.

From a relativistic point of view it would be more natural to discretize space and time right from the start, simultaneously and with the same scale ($\Delta_x=\Delta_t\ \textrm{finite}$). The resulting quantum simulation scheme would then take the form of a network of local quantum gates, homogeneously repeated across space and time---a Quantum Cellular Automata (QCA). \cite{SchumacherWerner,ArrighiUCAUSAL,ArrighiPQCA}. Feynman himself introduced QCA right along with the idea of quantum simulation \cite{FeynmanQCA}. He had pursued similar ideas earlier on in the one-particle sector, with his attractively simple `checkerboard model' of the electron in discrete $(1+1)$--spacetime \cite{Feynman_chessboard}. Later, the one-particle sector of QCA became known as Quantum Walks (QW), and was found to provide quantum simulation schemes for non-interacting fundamental particles \cite{BenziSucci,birula,meyer1996quantum,di2016quantum} in $(3+1)$--spacetime \cite{ArrighiDirac,marquez2017fermion}, be it curved \cite{di2013quantum,ArrighiGRDirac,ArrighiGRDirac3D, mallick2019simulating} or not, or in the presence of an electromagnetic field \cite{MolfettaDebbasch2014Curved, CGW18} or more in general a Yang-Mills interaction \cite{arnault2016quantum}. Some of these were implemented \cite{WernerElectricQW,Sciarrino}. The sense in which QW are Lorentz-covariant was made explicit \cite{ArrighiLORENTZ, PaviaLORENTZ, PaviaLORENTZ2, DebbaschLORENTZ}. The bounded speed of light is very natural in circuit-based quantum simulation, as it is directly enforced by the wiring between the local quantum gates. 

Yet, in spite of their many successes, discrete-spacetime models have fallen short of being able to account for realistic interacting QFT so far. The quantum simulation results over the multi-particle sector of QW (namely, QCA) are either abstract (e.g. universality \cite{ArrighiQGOL}) or phenomenological (e.g. molecular binding \cite{ahlbrecht2012molecular,PaviaMolecular}). An exception is a \cite{ArrighiToyQED}, where contact is made with $(1+1)$--QED in two ways : by mimicking its construction and by informally recovering its main phenomenology. All this points out to a core difficulty : there is no clear sense in which discrete-spacetime models of interacting particles have continuum limit ($\Delta_x=\Delta_t\longrightarrow 0$), in fact it is not even clear that interacting QFT themselves have such a continuum limit. In many ways, the classical Lagrangian that serves as departure point of a QFT is but a partial prescription for a numerical scheme (e.g. a regularized Feynman path integral), whose convergence is often challenging (renormalization). Continuous-spacetime does not seem to be the friendly place where QCA and QFT should meet. 

Clearly, Hamiltonian-based and QCA-based simulation schemes both have their advantages. It would be nice to have the best of both worlds: a discrete-spacetime model ($\Delta_x=\Delta_t$\ \textrm{finite}) that would be plastic enough to support both a non-relativistic continuous-time discrete-space limit ($\Delta_x$\ \textrm{finite}, $\Delta_t\longrightarrow 0$), in order to establish contact with the discrete-space continuous-time formulation of the QFT, and a fully relativistic spacetime-continuum limit ($\Delta_x=\Delta_t\longrightarrow 0$) limit. For a proof-of-concept we should aim for the free Dirac QFT first, and build a QW that converges both to its continuous-time discrete-space formulation (namely ``Lattice fermions'', i.e. the free part of the Kogut-Susskind Hamiltonian) and to its continuous-spacetime formulation (the Dirac equation). This is exactly what we achieve in this paper. 

For our construction, we needed `plasticity', in the sense of a tunable speed of propagation. Indeed intuitively, during the process where the continous-time limit discrete-space is taken, whenever $\Delta_t$ gets halved relative to $\Delta_x$, so is the particle's speed---because it gets half the time to propagate. This in turn is analogous to a change of coordinates, relabelling event $(t,x)$ into $(2t,x)$ in General Relativity. In order to keep physical distances the same, a synchronous metric $g=\textrm{diag}(1,-g_{xx})$ then becomes $g'=\textrm{diag}(1,-4 g_{xx})$ under such a change of coordinates. The original curved Dirac QW \cite{di2013quantum} is precisely able to handle any synchronous metric in the massless case; this was the starting point of our construction. Numerous trial--and--error modifications where needed, however, in order to control the relative scalings of $\Delta_t$ and $\Delta_x$ and in order to handle the mass elegantly. No wonder, therefore, that our result handles the case of curved $(1+1)$--spacetime `for free'. Our QW yields an original curved lattice-fermions Hamiltonian, never appeared in the literature \cite{villegas2015lattice,yamamoto2014lattice}, in the continuous-time discrete-space limit, and the standard curved Dirac equation in the spacetime-limit.

{\em Roadmap.} Section \ref{sec:themodel} presents the QW. Section \ref{sec:limits} shows the different limits it supports (see Fig \ref{FigUnifiedQW}). Section \ref{sec:curved} deals with synchronous curved $(1+1)$--spacetime. Section \ref{sec:qca} promotes the one-particle sector QW, to the many--non--interacting--particles sector, of a QCA. Section \ref{sec:conclusion} summarizes the results, and concludes.

\section{The model}\label{sec:themodel}

We consider a QW over the $(1+1)$--spacetime grid, which we refer to as the `Plastic QW'. Its coin or spin degree of freedom lies $\mathcal{H}_2$, for which we may chose some orthonormal basis $\{\ket{v^-}, \ket{v^+}\}$. The overall state of the walker lies the composite Hilbert space $\mathcal{H}_2\otimes \mathcal{H}_\mathbb{Z}$ and may be thus be written $\Psi=\sum_l \psi^+(l) \ket{v_+}\otimes\ket{l} + \psi^-(l) \ket{v_-}\otimes\ket{l}$, where the scalar field $\psi^+$ (resp. $\psi^-$) gives, at every position $m\in \mathbb{Z}$, the amplitude of the particle being there and about to move left (resp. right). We use $(j,l) \in \mathbb{N} \times \mathbb{Z}$, to label respectively instants and points in space and let:
\begin{equation}
\Psi_{j+2}=W \Psi_j
\label{eq:FDE0}
\end{equation}
where 
\begin{equation}
W = \Lambda^{-\kappa} \hat{S}({\hat{C}_{-\zeta}}\otimes \text{Id}_\mathbb{Z}) \hat{S}(\hat{C}_{\zeta}\otimes \text{Id}_\mathbb{Z} )\Lambda^\kappa
\end{equation}
with
$\hat{S}$ a state-dependent shift operator such that 
\begin{equation}
(\hat{S}\Psi)_{j,l}  =\begin{pmatrix}\psi^+_{j+1,l}\\\psi^-_{j-1,l}\end{pmatrix},
\end{equation} 

$\hat{C}_\zeta$ an element of $U(2)$ that depends on angles $\theta$ and $\zeta$,
\begin{equation}
\hat{C}_\zeta=   \begin{pmatrix}  -\cos \theta & e^{-i \zeta}\sin \theta  \\
e^{i \zeta} \sin \theta  & \cos \theta 
\end{pmatrix}
\end{equation}
and $\Lambda$ another element of $U(2)$ that depends on $f^\pm =\sqrt{1-c}\pm\sqrt{c+1}$,
\begin{equation}
\Lambda = \frac{1}{2}\left(
\begin{array}{cc}
- f^- &  f^+ \\
 f^+ &  f^- \\
\end{array}
\right).
\end{equation}
Later $c\in[0,1]$ will be interpreted as a speed of propagation or a hopping rate.

To investigate the continuous limit, we first introduce a time discretization step $\Delta_t$ and a space discretization step $\Delta_x$. We then introduce, for any parameter $a$ appearing in Eq. \eqref{eq:FDE0}, a field $\tilde a$ over the spacetime positions $\mathbb{R}^+ \times \mathbb{R}$, such that $a_{j,l}=\tilde a(t_j,x_l)$, with $t_j=j \Delta_t$, and $x_l = l \Delta_x$. Moreover the translation operator $\hat S$ will now proceed by $\Delta_x$ steps, so that 
$(S\widetilde{\Psi})(x_l)=\left(\widetilde{\psi}^+(x_l+\Delta_x),\widetilde{\psi}^-(x_l-\Delta_x)\right)^\top$ i.e. $S\widetilde{\Psi}=\exp( i \sigma_z \Delta_x p )\widetilde{\Psi}$, with $p=- i \partial_x$. 
Eq. \eqref{eq:FDE0} then reads: 
\begin{equation}
\widetilde{\Psi}(t_j+2\Delta_t) =  \widetilde{W}^{\theta}_{\zeta} \widetilde{\Psi}(t_j).
\label{eq:FDE2}
\end{equation}
Let us drop the tildes to lighten the notation. We suppose that all functions are $C^2$; for a more detailed analysis of the regularity condition which make these kinds of schemes convergent, the reader is referred to \cite{ArrighiDirac}.\\
Now, the continuum limit is the couple of differential equations obtained from Eq. \eqref{eq:FDE2} by letting both $\Delta_t$ and $\Delta_x$ go to zero. But how do we choose to let them go to zero, and what happens to the parameters, then? 
 
First, let us parametrize the time and space steps with a common $\varepsilon$:
\begin{equation}
\Delta_t = \varepsilon \hspace{1cm} \Delta_x = \varepsilon^{1-\alpha}\label{eq:scalings1}
\end{equation}
where $\alpha\in[0,1]$ will allow us to have $\Delta_t$ and $\Delta_x$ tend to zero differently.
Second, let us parametrize the positive real number $\kappa$, and the angles $\theta$, $\zeta$, by the same $\varepsilon$:
\begin{align}
\kappa&= \varepsilon^\alpha,\\
\theta &= \arccos(c \kappa),\\
\zeta & =  m \frac{(-1)^{\kappa}\varepsilon }{\sin(\theta)}.\\
\label{eq:scalings2}
\end{align}
Later $m\geq 0$ will be interpreted as a mass.\\
Summarizing,
\begin{itemize}
\item $\varepsilon$ will be taken to zero, triggering the continuum limit;
\item $\alpha$ will remain fixed throughout the limit, governing which type of continuum limit is being taken;
\item $m, c$ will remain fixed throughout the limit, stating which mass and speed are being simulated; 
\item $\Delta_t, \Delta_x, \kappa, \theta, \zeta$ will vary throughout the limit, entirely determined by the above four.
\end{itemize}

Notice that when we take the $\varepsilon\longrightarrow 0$ limit, $W$ remains unitary. For instance, focussing on the top-left entry of $C_\zeta$ we need $\cos \theta =c \varepsilon^\alpha \leq 1$, which requires that $\alpha \geq 0$ as already imposed.
Altogether, these jets define a family of QWs indexed by $\varepsilon$, whose embedding in spacetime, and defining angles, depend on $\varepsilon$. The continuum limit of Eq. \eqref{eq:FDE2} can then be investigated by Taylor expanding $\Psi(t,x)$ and $W^{\varepsilon,\alpha}$ for different values of $\alpha$.

\section{Continuum limit and scalings}\label{sec:limits}

Substituting for $\theta$ as in \eqref{eq:scalings2}, the coin operator reads
\begin{equation}
C^{\varepsilon,\alpha}_\zeta =  \big(-\varepsilon^\alpha c \sigma_z   + (\sigma_x \cos\zeta + \sigma_y \sin\zeta) \sqrt{1- c^2 \varepsilon^{2\alpha}} \big).
\label{eq:W0}
\end{equation}

\subparagraph{Case (i): $\mathbf{\alpha= 1.}$} In this case $\Delta_x = O(1)$ does not scale with $\varepsilon$ and so the translation operator is along fixed steps $\Delta_x$. In the leading order the operator $W^{\varepsilon,1}$ depends linearly on $\varepsilon$. 
The coin operator $C_\zeta$ reads:
\begin{equation}
C^{\varepsilon,1}_\zeta \simeq  - \varepsilon c \sigma_z + \sigma_x \cos(m \varepsilon)+ \sigma_y\sin(m \varepsilon)
) + O(\varepsilon^2).
\label{eq:W0C1}
\end{equation}
It is straightforward, after some simple algebra to derive the evolution operator at the zero-th order in $\varepsilon$:
\begin{equation}
W^{0,1} =  \text{Id}.
\label{eq:case1}
\end{equation}
Then the evolution operator reads:
\begin{equation}
W^{\varepsilon,1} \simeq \text{Id} - 2  \varepsilon ( i c \sigma_x\sin(i \Delta_x\partial_x)e^{\Delta_x \sigma_z \partial_x  } - i m \sigma_z)  + O(\varepsilon^2)
\label{eq:case1bis}
\end{equation}
Replacing this into Eq.\eqref{eq:FDE2}, and expanding $\Psi$ around $(x,t)$ we obtain:
\begin{small}
\begin{align*}
&\Psi + \varepsilon 2\partial_t \Psi\\
&= ( \text{Id} -  2  \varepsilon  i c \sigma_x\sin(i \Delta_x\partial_x)e^{\Delta_x \sigma_z \partial_x  } +2 i \varepsilon  m \sigma_z)\Psi + O(\varepsilon^2) \label{eq:casei}
\end{align*}
\end{small}

In the formal limit when $\varepsilon \longrightarrow 0$ this coincides with the hamiltonian equation:
\begin{equation}
i \partial_t  \Psi = H_L \Psi
\end{equation}
where 
\begin{equation}
H_L =  c \sigma_x\sin(i \Delta_x\partial_x)e^{\Delta_x \sigma_z \partial_x  } -   m \sigma_z.
\label{eq:case1eq} 
\end{equation}
This $H_L$ is precisely the Dirac Hamiltonian in the vacuum for a lattice fermion on a one dimensional grid (up to some minor re-encoding depending on conventions, see Sec. V for more details), i.e. the continuous-time discrete-space counterpart of the Dirac equation\footnote{Notice that $H_L$ commutes with $\sigma_x$, thus preserving the chiral symmetry w.r.t the components of the spinor $\Psi = (\psi^+,\psi^-)^\top$. This will no longer be true when we introduce the mass, which notoriously breaks chirality.}. 
Indeed the standard, Dirac equation in continuous spacetime, can be recovered at the level of \eqref{eq:case1eq} by setting $\Delta_x=\epsilon$ a posteriori, and computing the leading order of the expansion around $\epsilon=0$, which is
$i \partial_t \Psi  =( c \sigma_x \partial_x - m\sigma_z   )\Psi  + O(\epsilon^2)$, i.e. in the formal limit when $\epsilon \longrightarrow 0$,
\begin{align*}
i \partial_t  \Psi &= H_D \Psi\\
H_D &= c \sigma_x \partial_x -  m\sigma_z  
\end{align*}
Can we get to $H_D$ directly? 
 
\subparagraph{Case (ii): $0<\alpha<1$.} 
In this case the leading order of the translation operator is
\begin{equation}
 \hat{S}\simeq (\text{Id} + \varepsilon^{1-\alpha} \sigma_z \partial_x),
\end{equation}
whereas that of the coin operator is:
\begin{equation}
C^{\varepsilon,\alpha}_\zeta =  \big(-\varepsilon^\alpha c \sigma_z  + (\sigma_x \cos\zeta + \sigma_y \sin\zeta) \sqrt{1- c^2 \varepsilon^{2\alpha}} \big).
\label{eq:W0bis}
\end{equation}
The leading order of the Taylor expansion of the evolution operator reads:
\begin{equation}
W^{\varepsilon,\alpha} \simeq \text{Id} + 2 \varepsilon (-i c \sigma_x \partial_x + i m \sigma_z)+ O(\varepsilon^{1+\alpha}) 
\label{eq:case2}
\end{equation}
which directly recovers the standard, massive Dirac equation in continuous time :
\begin{equation}
i \partial_t \Psi = H_D \Psi.
\end{equation}
Notice that this result arises from the fact that the leading orders are given by terms of the kind $c \varepsilon^{\alpha} \varepsilon^{1-\alpha}\partial_x$, which no longer depend on $\alpha$, for a final result of order $O(\varepsilon)$.
Thus asking that $0 <\alpha < 1$, and thereby enforcing that $\Delta_t\longrightarrow 0$ faster than $\Delta_x\longrightarrow 0$, yields the same result than letting $\Delta_t\longrightarrow 0$, and then $\Delta_x\longrightarrow 0$, successively. Now, what if we let both of them go to zero at the same rate? 

\subparagraph{Case (iii): $\alpha=0$.} In this case the leading order of the translation operator is
\begin{equation}
 \hat{S} \simeq (\text{Id} +  \varepsilon \sigma_z  \partial_x) + O(\varepsilon^2),
\end{equation}
and the quantum coin becomes:
\begin{equation}
C^{\varepsilon,0}_\zeta =  -\sigma_z c + (\sigma_x \cos\zeta +\sigma_y \sin\zeta) \sqrt{1- c^2} .
\end{equation}
This special case is somehow opposite to that of Case (i), where the coin operator was scaling in $\varepsilon$, and the shift operator was independent of it. The leading order in $\varepsilon$ of the evolution operator leads to:
\begin{equation}
W^{\varepsilon,0} \simeq \text{Id}  + 2\Lambda \left( -i c \sigma_x \partial_x + i m \sigma_z) \right)\Lambda^{-1} + O(\varepsilon^{2}).
\label{eq:case3}
\end{equation}
Again the formal limit yields
\begin{align*}
i \partial_t \Psi &= H_D \Psi.\\
\end{align*}

Summarizing the results so far: 
\begin{theorem}\label{th:limits} Fix $m\geq 0$, $c\in [0,1]$. For different values of $\alpha\in [0,1]$, consider the family of QWs, parametrized by $\varepsilon\geq 0$:
\begin{equation}
\Psi(t_j+2\Delta_t)=W^{\varepsilon,\alpha}\Psi(t_j)
\label{eq:FDE}
\end{equation}
where 
\begin{equation}
W^{\varepsilon,\alpha} = \Lambda^{-\kappa} \hat{S}({\hat{C}_{-\zeta}}\otimes \text{Id}_\mathbb{Z}) \hat{S}(\hat{C}_{\zeta}\otimes \text{Id}_\mathbb{Z} )\Lambda^\kappa
\end{equation}
with
\begin{equation}
S=\exp(\sigma_z\Delta_x \partial_x),
\end{equation}
\begin{equation}
\hat{C}_{\zeta}=   \begin{pmatrix} - \cos \theta & e^{-i \zeta}\sin \theta  \\
e^{i \zeta} \sin \theta  & \cos \theta 
\end{pmatrix},
\end{equation}
\begin{equation}
\Lambda = \frac{1}{2}\left(
\begin{array}{cc}
- f^- &  f^+ \\
 f^+ &  f^- \\
\end{array}
\right), f^\pm =\sqrt{1-c}\pm\sqrt{c+1},
\end{equation}
\begin{align}
\kappa&= \varepsilon^\alpha,\\
\theta &= \arccos(c \kappa),\\
\zeta & =  m \frac{(-1)^{\kappa}\varepsilon }{\sin(\theta)}.
\end{align}
For any $0\leq\alpha\leq 1$, the $\varepsilon$--parametrized family admits a continuous time limit as $\varepsilon\longrightarrow 0$. For $\alpha =1$, this continuous-time limit is discrete in space. For $0\leq\alpha<1$ this continuous-time limit is also continuous in space. 
\begin{small}
\[ i \partial_t  \Psi = c \sigma_x\sin(i \Delta_x\partial_x)e^{\Delta_x \sigma_z \partial_x  }\Psi -   m \sigma_z\Psi , \hspace{1cm}  \text{for} \hspace{0.2cm} \alpha = 1.\] 
\[ i \partial_t \Psi = c \sigma_x \partial_x\Psi - m\sigma_z \Psi , \hspace{1.7cm} \text{for} \hspace{0.2cm}  0\leq \alpha <  1\] 
\end{small}
\label{theo1}
In both cases, the continuum limit is the differential equation corresponding to a massive Dirac fermion with mass $m$ and propagating at speed $c$. 
\end{theorem}

\section{Introducing a non homogeneous hopping rate $c$}\label{sec:curved}

The aim of this section is to generalize Th. \ref{theo1} to an inhomogeneous speed of propagation or hopping rate $c(t,x)$. In the continuous spacetime limit this corresponds to introducing a non-vanishing spacetime curvature. We will see that the spacetime-dependence of $c$ leads to a supplementary terms in the expansion of $W^{\varepsilon,\alpha}_{\zeta}$, proportional to $\Psi\partial_x C^{\varepsilon,\alpha}$ with $a = \{x, t\}$ . Let us look at each of the above case. 

Keeping the same scaling laws and dynamical equations as in Th. \ref{theo1}, we just need to generalise \eqref{eq:scalings2} as follows:
\begin{align}
\theta(t,x) &= \arccos(c(t,x) \kappa)\\
\zeta(t,x) & =  m \frac{(-1)^{\kappa}\varepsilon }{\sin(\theta(t,x))}
\label{eq:scalings2bis}
\end{align}
Again we assume that $c(t,x)$ is in $C^2$.

As in the previous section, we distinguish several $\varepsilon$--parametrized families of QWs, for different values of $\alpha$. 

\subparagraph{Case (i'):} $\alpha= 1$. The translation operator again no longer depends on $\varepsilon$. We have that
\begin{equation}
W^{\varepsilon,1}_{\zeta} = \text{Id} - 2 \varepsilon (\{e^{\Delta_x \sigma_z \partial_x} \sigma_x, e^{\Delta_x \sigma_z \partial_x} c(t,x)\sigma_z\} + i m \sigma_z)  + O(\varepsilon^2)
\end{equation} where \{,\} are the usual Poisson brackets. 
Replacing this once again into Eq.\eqref{eq:FDE2}, expanding $\Psi$ around $(x,t)$ and taking the formal limit for $\varepsilon \longrightarrow 0$, we recover the following hamiltonian equation:
\begin{align*}
i \partial_t  \Psi &= H_L(x,t) \Psi\\
H_L &=   -i \{e^{\Delta_x \sigma_z \partial_x} \sigma_x, e^{\Delta_x \sigma_z \partial_x} c(t,x)\sigma_z\} - m\sigma_z . 
\end{align*}


Quite surprisingly by setting $\Delta_x=\varepsilon$ a posteriori, we recover the curved massive Dirac equation in $(1+1)$-- dimensional spacetime:
\begin{equation}
i \partial_t \Psi =  \sigma_x\partial_x \Psi  + \frac{\sigma_x}{2}\Psi \partial_x c - m \sigma_z \Psi
\end{equation}
This suggests us that \textbf{Case (i')} may be a simple way to simulate the Dirac equation in a curved spacetime by the implementation of a continuous-time discrete-space QW, which to the best of our knowledge is an original result . 

Again we can get to the same PDE directly with by setting $0\geq\alpha<1$. Indeed, \textbf{Case (ii')} and \textbf{Case (iii')} are analogous to the homogeneous case, except for a supplementary term in the PDE represented by the spatial derivative of the coin $W^{\varepsilon,\alpha}_{m} \simeq W_m^{\varepsilon,\alpha} + \frac{1}{2} \partial_x W_{0} ^{\varepsilon,\alpha}$.  It is tedious but straightforward verify that Th. \ref{theo1} generalises as follows: 

\begin{theorem} \label{th:curvedlimits} 
Fix $m\geq 0$, $c\in [0,1]$. For different values of $\alpha\in [0,1]$, consider the family of QWs, parametrized by $\varepsilon\geq 0$ :
\begin{equation}
\Psi(t_j+2\Delta_t)=W^{\varepsilon,\alpha}\Psi(t_j)
\label{eq:FDE3}
\end{equation}
where 
\begin{equation}
W^{\varepsilon,\alpha} = \Lambda^{-\kappa} \hat{S}({\hat{C}_{-\zeta}}\otimes \text{Id}_\mathbb{Z}) \hat{S}(\hat{C}_{\zeta}\otimes \text{Id}_\mathbb{Z} )\Lambda^\kappa
\end{equation}
with
\begin{equation}
S=\exp(\sigma_z\Delta_x \partial_x),
\end{equation}
\begin{equation}
\hat{C}_{\zeta}=   \begin{pmatrix} - \cos \theta & e^{-i \zeta}\sin \theta  \\
e^{i \zeta} \sin \theta  & \cos \theta 
\end{pmatrix},
\end{equation}
\begin{equation}
\Lambda = \frac{1}{2}\left(
\begin{array}{cc}
- f^- &  f^+ \\
 f^+ &  f^- \\
\end{array}
\right), f^\pm =\sqrt{1-c}\pm\sqrt{c+1},
\end{equation}
\begin{align}
\kappa&= \varepsilon^\alpha,\\
\theta(t,x) &= \arccos(c(t,x) \kappa),\\
\zeta(t,x) & =  m \frac{(-1)^{\kappa}\varepsilon }{\sin(\theta(t,x))}.
\end{align}
For any $0\leq\alpha\leq 1$, the $\varepsilon$--parametrized family admits a continuous time limit as $\varepsilon\longrightarrow 0$. For $\alpha =1$, this continuous-time limit is discrete in space. For $0\leq\alpha<1$ this continuous-time limit is also continuous in space. 
\begin{small}
\[ i \partial_t  \Psi = -i \{e^{\Delta_x \sigma_z \partial_x} \sigma_x, e^{\Delta_x \sigma_z \partial_x} c(t,x)\sigma_z\} - m\sigma_z , \hspace{0.55cm}  \text{for} \hspace{0.2cm} \alpha = 1\] 
\[ i \partial_t \Psi = c(t,x) \sigma_x \partial_x \Psi +  \sigma_x  \Psi \frac{1}{2}\partial_x c(t,x) - m \sigma_z \Psi, \hspace{0.5cm} \text{for} \hspace{0.2cm}  0\leq \alpha <  1\] 
\end{small}
\label{theo2}
In both cases, the continuum limit is the differential equation for a massive Dirac fermion propagating on an arbitrary in curved spacetime, with synchronous coordinates---i.e. coordinates in which the metric tensor has coefficients $g^{00}=1$, $g^{01}=g^{10}=0$ and $g^{11}=-\frac{1}{c(x,t)^2}$.
\end{theorem}

\section{Many-particle model}\label{sec:qca}

Let us now extend our formalism to the multi-particle sector. We will construct a `Plastic' Quantum Cellular Automata in $(1+1)$--dimensions aiming to model Dirac fermions without interaction. It could be argued that the Plastic QCA will be nothing but many Plastic QW weaved together, which is true of course. Still this implies a not so obvious shift in mathematical formalism, and constitutes a mandatory prior step in order to later approach the modelling of interacting QFTs.\\

\begin{figure}
{\centering
\includegraphics[scale=0.3]{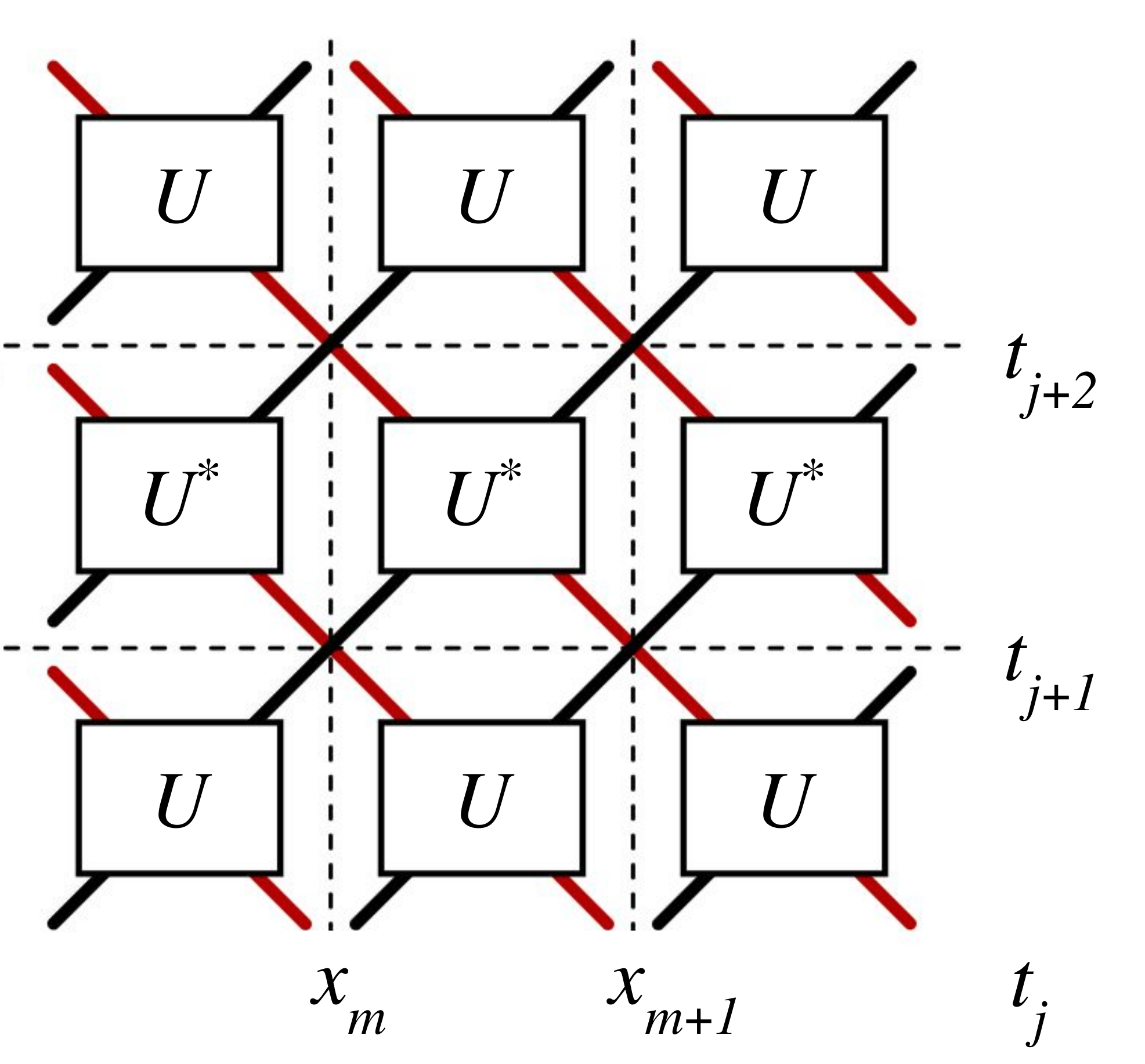}
\caption{{\em The Plastic QCA.}\label{fig:QCA} A layer of $U$ is a applied, alternated with a swap, then a layer of $U^*$, and a swap again.}
}
\end{figure}

For this Plastic QCA we adopt the conventions depicted in Fig.~\ref{fig:QCA}. Each red (resp. black) wire carries a qubit, which codes for the presence or absence of a left-moving (resp. right-moving) electron. The cells are located right at the crossings (or right afterwards). Therefore their Hilbert space is ${\cal H}={\cal H}_L\otimes{\cal H}_R$ with ${\cal H}_L={\cal H}_R=\mathbb{C}^2$, i.e. they are made out of two subcells, accounting the left-moving and right-moving modes respectively.\\
The dynamics acts in four steps. The even steps act according to a partition $\mathcal{P}$, the odd step act according to the same partition, but half-shifted, i.e. $\mathcal{Q}$. By `partition', we mean a way of grouping subcells, to then act upon them with the synchronous application of a local unitary, across space. Partition $\mathcal{P}$ groups the right subcell of the left cells, with the left subcell of the right cells, in order to act upon them with a local unitary gate. In the first step, this local unitary gate is $U$, which can therefore be thought of as being located at positions $(t_j+\frac{\Delta_t}{2},x_l+\frac{\Delta_x}{2})$ for $j\in 2\mathbb{N}$ and $l\in\mathbb{Z}$. In the third step, this local unitary gate is $U^*$, which can therefore be thought of as being located at positions $(t_j+\frac{3\Delta_t}{2},x_l+\frac{\Delta_x}{2})$ for $j\in 2\mathbb{N}+1$ and $m\in\mathbb{Z}$. Partition $\mathcal{Q}$ simply regroups the left and right subcell of a given cell, i.e. the local unitary acts per cell. In both the second and fourth steps the local unitary is $V$, which can therefore be thought of as being located at positions $(t_j,x_l)$ for $j\in \mathbb{N}$ and $l\in\mathbb{Z}$.
The local unitary gate $V$ is the simplest:
\begin{equation}
V=\begin{pmatrix}1 & 0 & 0 & 0\\
0 & 0&  1& 0\\
0 & 1&  0& 0\\
0 & 0 & 0& -1
\end{pmatrix}
\end{equation}
\begin{small}
\begin{align}
= |00\rangle\langle 00|+|01\rangle\langle 10|\\
+|10\rangle\langle 01|-|11\rangle\langle 11|,
\end{align}
\end{small}
i.e. it is a mere swap, just represented by the wire crossings in Fig.~\ref{fig:QCA}. Its only oddity is the $(-1)$ phase when two excitations permute: we refer to \cite{ArrighiToyQED} for a justification.
The local unitary gate $U$, on the other hand, acts non-trivially in the one-particle sector, in a way which is related to the coin operator:
\begin{equation}
U=\begin{pmatrix}1 & 0 & 0 & 0\\
0 & e^{-i \zeta}\sin\theta &  -\cos\theta & 0\\
0 & \cos\theta & e^{i \zeta}\sin\theta & 0\\
0 & 0 & 0 & -1
\end{pmatrix}
\end{equation}
\begin{small}
\begin{align}
= |00\rangle\langle 00|+ e^{-i \zeta}\sin\theta  |01\rangle\langle 01|   -\cos\theta   |01\rangle\langle 10|\\+ \cos\theta |10\rangle\langle 01|
+e^{i \zeta}\sin\theta  |10\rangle\langle 10|-  |11\rangle\langle 11|.
\end{align}
\end{small}
Altogether, the global evolution is:
\[
\Psi(t_j+2\Delta_t)= G \Psi(t_j)
\] 
\[
G= \bigotimes_{\cal Q} V \bigotimes_{{\cal P}} U^* \bigotimes_{\cal Q} V \bigotimes_{{\cal P}} U
\] 

We wish to argue that in the discrete-space continuous-time limit, this Plastic QCA behaves like many-particle non-interacting lattice fermions. We will do so by focussing on the one-particle sector behaviour of this QCA, and then second quantizing it by standard methods. Here is the one-particle sector:
\begin{small}
\begin{align}
\Psi(t_j) =\sum_{l\in\mathbb{Z}} &\psi^+(x_l,t_j) \ldots\ket{00}^{l-1}\ket{10}^l\ket{00}^{l+1}\ldots+\\
&\psi^-(x_l,t_j) \ldots\ket{00}^{l-1}\ket{01}^l\ket{00}^{l+1}\ldots
\end{align}
\end{small}

After a tedious but straightforward calculation, we can extract the recurrence relation for the amplitudes $\psi_l(t_j) = \{\psi^+(x_l,t_j),\psi^-(x_l,t_j)\}$ at time $t_j+2\Delta_t$; Taylor expand them around $\varepsilon = 0$ using the scalings \eqref{eq:scalings2} for $\alpha=1$; and take the formal limit for $\varepsilon \longrightarrow 0$. This yields, with a spacetime dependent $U$, to the discretized one-particle Hamiltonian:


\begin{small}
\begin{multline}
H_{\mathcal{G}d} \psi_l(t) =\frac{i}{2 } \sigma_x \left(c_{l-\frac{\Delta_x}{2}}\psi_{l-\Delta_x}-c_{l+\frac{\Delta_x}{2}}\psi_{l+\Delta_x}\right) - m \sigma_z \psi_l.
\label{eq:discretDH_C}
\end{multline}
\end{small}
This is our proposed, curved spacetime lattice fermions Hamiltonian. One can check that in the continuous-space limit, this Hamiltonian goes to the standard, continuous spacetime Hamiltonian of the curved $(1+1)$--spacetime Dirac equation in synchronous coordinates. Moreover, the proposed Hamiltonian \eqref{eq:discretDH_C} coincides, in the case of a homogeneous and static metric, with your usual one-particle lattice fermion Hamiltonian:

\begin{equation}
H_{d} \psi_l(t) =\frac{i c}{2 } \sigma_x \left(\psi_{l-\Delta_x}-\psi_{l+\Delta_x}\right) - m \sigma_z \psi_l.
\label{eq:discretDH}
\end{equation}

This latter Hamiltonian, when massless, commutes with $\sigma_x$ (preserving the chiral symmetry). Any other massless representation, e.g. that one commuting with $\sigma_y$, can be recovered by choosing different phase in the operator $U$\footnote{In order to recover a massless Hamiltonian commuting with $\sigma_y$ we can choose the operator \begin{equation}
U'=\begin{pmatrix}1 & 0 & 0 & 0\\
0 & e^{-i \zeta}\sin \theta &  -\cos \theta & 0\\
0 & \cos \theta & -e^{i \zeta}\sin \theta & 0\\
0 & 0 & 0 & -1
\end{pmatrix}
\end{equation}}.

The one-particle lattice fermion Hamiltonian of \eqref{eq:discretDH} is quite comparable to that of \eqref{eq:case1eq}, but has the advantage of being rather well-known and admitting a standard second-quantization \cite{kogut1975hamiltonian}. Moreover the two can be related in discrete spacetime picture in the sence that the one-particle sector of $G$ coincides with $W$ up to an initial encoding. Indeed, $G$ acts in the one-particle sector as follows: 
\begin{equation}
W' =( \hat{S}^-{\hat{C}_{-\zeta}} \hat{S}^+) ( \hat{S}^-{\hat{C}_{\zeta}} \hat{S}^+ )
\end{equation}
where $\hat{S}^\pm$ are partial shifts in space:
$$
(\hat{S}^+\Psi)_{j,l}  =\begin{pmatrix}\psi^+_{j,l+1}\\\psi^-_{j,l}\end{pmatrix}
\quad 
(\hat{S}^-\Psi)_{j,l}  =\begin{pmatrix}\psi^+_{j,l}\\\psi^-_{j,l-1}\end{pmatrix}
$$
The operator is thus equivalent to $W^\theta_\zeta$ up to an encoding $E = {S^+}$ and setting $\Lambda= I_2$:
\begin{equation}
{W'}^{\theta}_{\zeta} = E^\dagger{W}^{\theta}_{\zeta} E.
\end{equation}

From \eqref{eq:discretDH} we can derive the continuous-time the Kogut-Susskind Hamiltonian of the free Dirac QFT (a self-adjoint operator in the Fock space) using the more abstract language of modern quantum field theory. We can use the discretized single-particle Hamiltonian as a self-adjoint operator in the Fock space $\mathcal{F}$ according to
\begin{multline}
\hat{H} = \sum_l \hat{\Psi}^*(H_df_l)\hat\Psi(f_l) = \\ 
 \frac{1}{2} \sum_l\big[ \hat\Psi^{2\dagger}_{l+1} \hat\Psi^{1}_{l} + \hat\Psi^{2\dagger}_{l} \hat\Psi^{1}_{l+1} + h.c.\big] + m \sum_l \big[ \hat\Psi^{1\dagger}_{l} \hat\Psi^{1}_{l} - \hat\Psi^{2\dagger}_{l} \hat\Psi^{2}_{l}. \big] \label{eq:manyparticleKG}
\end{multline}
where the second quantized discretized Dirac field operator 
\begin{equation}
\hat\Psi^\alpha_l = \hat\Psi^\alpha_l(f_l) = \int dx \hat\Psi^\alpha(x) f^*_l(x)  \hspace{0.5cm} \alpha = \{+, -\}
\end{equation}
satisfies the above anti-commutation relations and the orthonormal set of basis functions $f_l$ spans the discretized Hilbert space $\mathbb{Z}$. Notice that the above equation takes the form of the Fermi-Hubbard equation where first term implements the hopping between the neighboring sites and the second term is a local term accounting the mass-term.

Altogether, this provides strong evidence that the Plactic QCA $G$ has discrete-space continuous-time limit the free Dirac Kogut-Susskind Hamiltonian \eqref{eq:manyparticleKG}. However, it would prove this directly in the many-particle sector, without restricting to the one-particle sector and then second-quantizing. We leave this question open.


\section{Conclusion}\label{sec:conclusion}

{\em Summary of results.} We introduced a Quantum Walk (QW) over the $(1+1)$--spacetime grid, given by \eqref{eq:FDE0}.
The QW has parameters $m$, $c$, $\varepsilon$, $\alpha$. The first two are parameters of the physics that we simulate : $m$ is the mass of the Dirac fermion, $c$ is the speed of light. The second two control the scaling of the discretization : $\varepsilon\longrightarrow 0$ whenever we take a continuum limit, whereas $0\geq\alpha\geq 1$ will remain fixed but determine which limit is to be taken, by specifying the relative scalings of $\Delta_t=\varepsilon$ and $\Delta_x=\varepsilon^{1-\alpha}$. When $\alpha=0$ the continuous-spacetime limit ($\Delta_x=\Delta_t\longrightarrow 0$) yields the Dirac equation. The same is true of all intermediate cases $0\leq\alpha<1$. But when $\alpha=1$, the continuous-time discrete-space limit ($\Delta_x  \textrm{finite},\Delta_t\longrightarrow 0$) gets triggered, and yields the lattice fermion Hamiltonian. The result is encapsulated in Th. \ref{th:limits}, and generalized to a spacetime dependent $c(x,t)$, in Th. \ref{th:curvedlimits} recovering the curved spacetime the lattice fermion and Dirac equation for synchronous coordinates. Finally the QW is made into a QCA by considering many non-interacting particles; in the limits this yields free Dirac QFT as expected, i.e. lattice fermions for many-non-interactive particles. 

{\em Perspectives.} The QCA may be viewed as a discrete-spacetime version of free Dirac QFT. In the same way that free Dirac QFT has as asymetric space-discretization the lattice fermions, this QCA has as asymetric continuous-time limit the lattice fermions. It is unitary and strictly respects the speed of light. Our long term aim is to add interactions this QCA, thereby obtaining a discrete-spacetime version of some interacting QFT in the style of \cite{ArrighiToyQED}---except that it would support a continuous-time limit towards the non-relativistic, discrete-space continuous-time reformulations of the interacting QFT, for due validation. Discrete-space continuous-time may well be the friendly place where QCA and QFT should meet, after all.\\ An unexpected by-product of this work is the provision of a space-discretization of the curved $(1+1)$--spacetime Dirac equation in synchronous coordinates, i.e. the provision of an original curved lattice fermions Hamiltonian. This opens the route for elaborating curved Kogut-Susskind Hamiltonian, and to eventually suggest Hamiltonian-based quantum simulators of interacting particles over a curved background.\\
An interesting remark raised by one of the anonymous referees is the following. On the one hand, the non-relativistic, naive lattice fermions Hamiltonians are known \cite{kogut1975hamiltonian} to suffer the fermion-doubling problem, i.e. a spurious degree of freedom. On the other hand, the Dirac QW does not suffer this problem \cite{ArrighiChiral}. An intriguing question is whether the Plastic QW hereby presented, which borrows from both worlds, suffers this problem or not. We leave this as an open question.

Finally, we wonder whether the Shr\"odinger limit ($c \longrightarrow \infty$) could be fitted into the picture.

\section{Competing interests}
The authors declare that there are no competing interests.

\section{Author Contribution}
GDM and PA contributed equally to the main results of the manuscript.

\section{Data Availability}
No data sets were generated or analysed during the current study.

\section*{Acknowledgements} The authors would like to thank Pablo Arnault and C\'edric B\'eny for motivating discussions.

\bibliographystyle{plain}	
\bibliography{biblio}

\end{document}